\documentclass[12pt]{iopart}
\usepackage{iopams}
\usepackage{color}
\usepackage{graphicx}
\usepackage{cite}

\begin{document}

\title[Exactly solved spin-1/2 Heisenberg-Ising ladder]{
Quantum phase transitions
in the exactly solved spin-1/2 Heisenberg-Ising ladder}
\author{Taras Verkholyak$^{1,2}$,
        Jozef Stre\v{c}ka$^{2}$}
\address{$^1$Institute for Condensed Matter Physics,
             National Academy of Sciences of Ukraine,
             1 Svientsitskii Street, L'viv-11, 79011, Ukraine}
\address{$^2$Department of Theoretical Physics and Astrophysics,
             Institute of Physics, P. J. \v{S}af\'{a}rik University,
             Park Angelinum 9, 040 01 Ko\v{s}ice, Slovak Republic}
\ead{werch@icmp.lviv.ua}


\begin{abstract}
Ground-state behaviour of the frustrated quantum spin-$\frac{1}{2}$ two-leg ladder
with the Heisenberg intra-rung and Ising inter-rung interactions is examined
in detail. The investigated model is transformed to the quantum Ising chain
with composite spins in an effective transverse and longitudinal field
by employing either the bond-state representation or the unitary transformation.
It is shown that the ground state of the Heisenberg-Ising ladder can be
descended from three exactly solvable models: the quantum
Ising chain in a transverse field, the 'classical' Ising chain in a longitudinal field
or the spin-chain model in a staggered longitudinal-transverse field. The last model serves
in evidence of the staggered bond phase with alternating singlet and triplet
bonds on the rungs of two-leg ladder, which appears at moderate values of the external magnetic
field and consequently leads to a fractional plateau at a half of the saturation magnetization.
The ground-state phase diagram totally consists of five ordered and one quantum paramagnetic phase, which are
separated from each other either by the lines of discontinuous or continuous quantum
phase transitions. The order parameters are exactly calculated for all five ordered phases
and the quantum paramagnetic phase is characterized through different short-range spin-spin correlations.
\end{abstract}


\submitto{\JPA}
\maketitle



\section{Introduction}

Quantum spin models with competing interactions represent quite
interesting and challenging topic for the current research
\cite{solstsci164,lnp645}. Such models show many unusual features
in their ground-state properties and are very sensitive to the
approximative schemes applied to them. An existence of exact
solutions is therefore quite important, since they provide
a rigorous information about the complex behaviour of frustrated models.
However, exact results for such systems are still limited (see \cite{miyahara2011}
and references cited therein).
The simplest examples of rigorously solved quantum spin models
are the models with the known dimerized \cite{miyahara2011,kumar2002,schmidt2005,gelle2008}
or even trimerized \cite{schmidt2010} ground states. Another example represents
frustrated quantum spin antiferromagnets in high magnetic fields
with the so-called localized-magnon eigenstates (see e.g. \cite{derzhko2007}
for recent review). It should be nevertheless noticed that the
aforementioned exact results are usually derived only under certain constraint
laid on the interaction parameters and those models are not
tractable quite generally.

On the other hand, the exact solutions for some spin-$\frac{1}{2}$
quantum spin ladders with multispin interaction allows to find, besides
the ground state, also thermodynamic properties quite rigorously \cite{barry1998,batchelor2007}.
Exactly solvable models, which admit the inclusion of frustration,
can also be constructed from the Ising models where the
decorations of quantum spins are included \cite{strecka2010,canova2006}.
The decoration-iteration procedure allows one to calculate exactly all
thermodynamic properties of the decorated models when the exact solution
for the corresponding Ising model is known. The distinctive feature
of these models is that the quantum decorations are separated from each other,
so that the Hamiltonian can be decomposed into the sum of commuting parts.
The eigenstates of the total Hamiltonian are then simply factorized
into a product of the eigenstates of its commuting parts.

New findings of the exactly solvable frustrated quantum spin models with
non-trivial (not simply factorizable) ground states are thus highly desirable.
The present article deals with the frustrated quantum spin-$\frac{1}{2}$
two-leg ladder with the Heisenberg intra-rung interaction and the Ising inter-rung
interactions between nearest-neighbouring spins from either the same or different leg.
Such a model can be regarded as an extension of the spin-$\frac{1}{2}$ Heisenberg-Ising bond alternating chain,
which was proposed and rigorously solved by Lieb, Schultz and Mattis \cite{lsm}.
Alternatively, this model can also be viewed as the generalization of the exactly solved
quantum compass ladder \cite{brze,brze10a,brze10b} when considering the Heisenberg rather than XZ-type
intra-rung interaction and accounting for the additional frustrating Ising inter-rung interaction.
Moreover, it is quite plausible to suspect that the exact results presented hereafter
for the spin-$\frac{1}{2}$ Heisenberg-Ising two-leg ladder may also bring insight into
the relevant behaviour of the corresponding Heisenberg two-leg ladder, which represents
a quite challenging and complex research problem in its own right \cite{rice,dago}.
The frustrated spin-$\frac{1}{2}$ Heisenberg two-leg ladder have been extensively
investigated by employing various independent numerical and analytical methods such as
density-matrix renormalization group \cite{legeza1997a,legeza1997b,wang2000,honecker2000},
numerical diagonalization \cite{honecker2000,okazaki2000,sakai2000,chandra2006,weihong1998}, series expansion \cite{weihong1998},
bosonization technique \cite{totsuka1998,kim1999,allen2000}, strong- and weak-coupling analysis \cite{mila1998,kim2008,starykh2004,hikihara2010,michaud2010},
valence-bond spin-wave theory \cite{xian1995}, variational matrix-product approach \cite{brehmer1998,kolezhuk1998},
and bond mean-field theory \cite{ramakko2007,azzouz2008,ramakko2008}. Among the most interesting results obtained for
this quantum spin chain, one could mention an existence of the columnar-dimer phase discussed in \cite{starykh2004,hikihara2010}
or a presence of the fractional plateau in the magnetization process examined in \cite{honecker2000,okazaki2000,sakai2000,chandra2006,michaud2010}.

The theoretical investigation of two-leg ladder models is motivated not only from
the academic point of view, but also from the experimental viewpoint, because
the two-leg ladder motif captures the magnetic structure of a certain class of real quasi-one-dimensional
magnetic materials. The most widespread family of two-leg ladder compounds form cuprates,
in which one finds both experimental representatives with the dominating intra-rung interaction
like SrCu$_2$O$_3$ \cite{azuma91,azuma94}, Cu$_2$(C$_5$H$_{12}$N$_2$)$_2$Cl$_4$ \cite{chiari,chabo97a,chabo97b,chabo98},
(C$_5$H$_{12}$N)$_2$CuBr$_4$ \cite{watson}, (5IAP)$_2$CuBr$_4$ \cite{landee}
as well as, the magnetic compounds with the dominating intra-leg interaction such as
KCuCl$_3$ \cite{shira,osawa02}, TlCuCl$_3$ \cite{osawa01,osawa03,matsu}, NH$_4$CuCl$_3$, KCuBr$_3$.
Understanding the low-temperature magnetism of two-leg ladder models also turns out to be
crucial for an explanation of the mechanism, which is responsible for the high-temperature
superconductivity of cuprates \cite{mae}. Besides the cuprates, another experimental representatives
of the two-leg ladder compounds represent vanadates (VO)$_2$P$_2$O$_7$ \cite{leo}, CaV$_2$O$_5$ and
MgV$_2$O$_5$ \cite{korotin}, as well as, the polyradical BIP-BNO \cite{katoh,hoso}.

The outline of the paper is as follows.
In section~\ref{model} the model is defined and the pseudospin representation is considered.
The ground-state properties of the model with and without external field are studied in section~\ref{ground_state}.
The most important findings are summarized in section~\ref{conclusions}.

\section{Heisenberg-Ising two-leg ladder}
\label{model}
\begin{figure}[b]
 \begin{center}
   \includegraphics[width=0.8\textwidth]{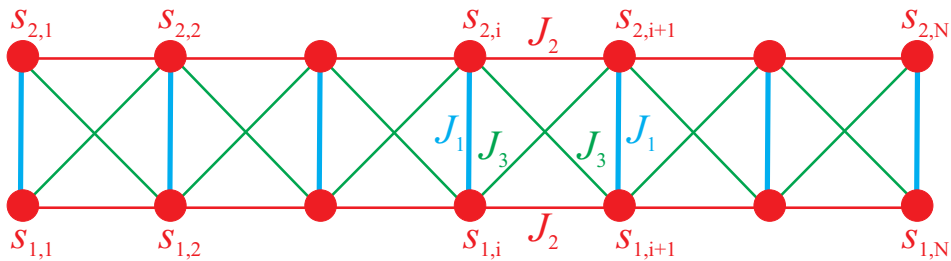}
 \end{center}
\caption{Quantum spin-$\frac{1}{2}$ Heisenberg-Ising two-leg ladder.
Thick (thin) lines denote the Heisenberg (Ising) bonds.}
\label{2leg_ladder}
\end{figure}

Let us define the spin-$\frac{1}{2}$ Heisenberg-Ising ladder through
the following Hamiltonian:
\begin{eqnarray}
\label{gen_ham1}
H&=&\sum_{i=1}^{N} \Big[ J_1({\mathbf s}_{1,i} \cdot {\mathbf s}_{2,i})_{\Delta}
+J_2(s_{1,i}^z s_{1,i+1}^z+s_{2,i}^z s_{2,i+1}^z)
\nonumber\\
&&+J_3 (s_{2,i}^z s_{1,i+1}^z+ s_{1,i}^z s_{2,i+1}^z)
-h(s_{1,i}^z+s_{2,i}^z) \Big],
\end{eqnarray}
where
$({\mathbf s}_{1,i} \cdot {\mathbf s}_{2,i})_{\Delta}
=s_{1,i}^x s_{2,i}^x + s_{1,i}^y s_{2,i}^y + \Delta s_{1,i}^z s_{2,i}^z$,
$s_{l,i}^\alpha$ are three spatial projections of spin-$\frac{1}{2}$ operator,
the first index denotes the number of leg, the second enumerates the site,
$J_1$ is the $XXZ$ Heisenberg intra-rung interaction between nearest-neighbour spins from the same rung,
$J_2$ is the Ising intra-leg interaction between nearest-neighbour spins from the same leg,
$J_3$ is the crossing (diagonal) Ising inter-rung interaction between next-nearest-neighbour spins from different rungs,
$h$ is the external magnetic field.
We also imply the periodic boundary conditions
${\mathbf s}_{1,N+1} \equiv {\mathbf s}_{1,1}$ and
${\mathbf s}_{2,N+1} \equiv {\mathbf s}_{2,1}$ along legs.
The coupling constants $J_2$ and $J_3$ can be interchanged by renumbering of the sites, as well as
their signs can be simultaneously reverted by spin rotations.
Therefore, the Hamiltonians $H(J_2,J_3)$, $H(J_3,J_2)$ and $H(-J_2,-J_3)$ in zero field
have equal eigenvalues and the corresponding models are thermodynamically equivalent.

It can be checked that $z$-projection of total spin on a rung $S_i^z=s_{1,i}^z+s_{2,i}^z$
commutes with the total Hamiltonian $[S_i^z,H]=0$ and hence, it represents
a conserved quantity. For further convenience, it is therefore advisable
to take advantage of the bond representation, which has been originally suggested
by Lieb, Schultz and Mattis for the Heisenberg-Ising chain \cite{lsm}.
Let us introduce the bond-state basis consisting of four state vectors:
\begin{eqnarray}
\fl && |\phi_{0,0}^{i}\rangle=\frac{1}{\sqrt2}
(|\! \downarrow_{1,i}\uparrow_{2,i}\rangle - |\! \uparrow_{1,i}\downarrow_{2,i}\rangle),
\; \;
|\phi_{1,0}^{i}\rangle=\frac{1}{\sqrt2}
(|\! \downarrow_{1,i}\uparrow_{2,i}\rangle + |\! \uparrow_{1,i}\downarrow_{2,i}\rangle),
\nonumber\\
\fl && |\phi_{1,-}^{i}\rangle=\frac{1}{\sqrt2}
(|\! \uparrow_{1,i}\uparrow_{2,i}\rangle - |\! \downarrow_{1,i}\downarrow_{2,i}\rangle),
\; \;
|\phi_{1,+}^{i}\rangle=\frac{1}{\sqrt2}
(|\! \uparrow_{1,i}\uparrow_{2,i}\rangle + |\! \downarrow_{1,i}\downarrow_{2,i}\rangle).
\end{eqnarray}
These states are the eigenstates of the Heisenberg coupling between two spins located
on the $i$th rung, i.e.
$|\phi_{0,0}^{i}\rangle$ is the singlet-bond state,
$|\phi_{1,0}^{i}\rangle$, $|\phi_{1,\pm}^{i}\rangle$ are triplet states.
Following \cite{lsm} two subspaces can be singled out in the bond space:
i) $|\phi_{0,0}^{i}\rangle$, $|\phi_{1,0}^{i}\rangle$
and ii) $|\phi_{1,+}^{i}\rangle$, $|\phi_{1,-}^{i}\rangle$.
The first and second subspaces corresponds to $(S_i^z)^2=0$ and $(S_i^z)^2=1$ respectively,
and the Hamiltonian can be diagonalized separately in each subspace.
We can introduce the index of subspace at $i$-th site:
$n_i=0 (1)$ if the given state is in the subspace
with $(S_i^z)^2=0$ ($(S_i^z)^2=1$) spanned by states
$|\phi_{0,0}^{i}\rangle$, $|\phi_{1,0}^{i}\rangle$
($|\phi_{1,-}^{i}\rangle$, $|\phi_{1,+}^{i}\rangle$).
According to \cite{lsm}, let us also call the bond states as purity (impurity) states
if $n_i=0(1)$.
If one makes pseudospin notations for the states
$|\phi_{0,0}^{i}\rangle=|\! \! \! \downarrow\rangle_0^i$, $|\phi_{1,0}^{i}\rangle=|\! \! \! \uparrow\rangle_0^i$,
the action of the spin-$\frac{1}{2}$ operators in the subspace $n_i=0$ can be
expressed in terms of new raising and lowering operators:
\begin{eqnarray}
\label{new_op0}
&& s_{1,i}^z = -\frac{1}{2}(a_i^+ + a_i)=-\tilde{s}_i^x, \; \; \;
s_{2,i}^z = \frac{1}{2}(a_i^+ + a_i)=\tilde{s}_i^x,  \; \; \;
\nonumber\\
&&
({\mathbf s}_{1,i} \cdot {\mathbf s}_{2,i})_{\Delta} = a_i^+ a_i -\frac{2+\Delta}{4}
=\tilde{s}_i^z-\frac{\Delta}{4}.
\end{eqnarray}
The operators $a_i^+$, $a_j$ satisfy the Pauli algebra
($\{a_i,a_i^+\}=0$, $\{a_i,a_i\}=\{a_i^+,a_i^+\}=0$, $[a_i,a_j]=[a_i,a_j^+]=[a_i^+,a_j^+]=0$ for $i\neq j$),
and half of their sum can be identified as a new pseudospin operator $\tilde{s}_i^x$.

Analogously, one can consider the pseudospin representation of the subspace
$n_i=1$ ($|\phi_{1,-}^{i}\rangle=|\! \! \downarrow\rangle_1^i$, $|\phi_{1,+}^{i}\rangle=|\! \! \uparrow\rangle_1^i$)
and find the action of spin operators in it as follows:
\begin{eqnarray}
\label{new_op1}
\fl && s_{1,i}^z = \frac{1}{2}(a_i^+ + a_i)=\tilde{s}_i^x, \; \;
s_{2,i}^z = \frac{1}{2}(a_i^+ + a_i)=\tilde{s}_i^x,  \; \;
({\mathbf s}_{1,i} \cdot {\mathbf s}_{2,i})_{\Delta} = \frac{\Delta}{4}.
\end{eqnarray}
Combining (\ref{new_op0}), (\ref{new_op1})
we find the general expressions for the pseudospin representation of these operators,
which are valid in both subspaces:
\begin{eqnarray}
\label{new_op}
\fl && s_{1,i}^z = (2n_i-1)\tilde{s}_i^x, \; \;
    s_{2,i}^z = \tilde{s}_i^x, \; \;
({\mathbf s}_{1,i} \cdot {\mathbf s}_{2,i})_{\Delta} =(1-n_i)\tilde{s}_i^z+\frac{\Delta}{4}(2n_i-1).
\end{eqnarray}
The effective Hamiltonian can be rewritten in terms of new operators as follows:
\begin{eqnarray}
\label{gen_ham2}
 H&=&\sum_{i=1}^{N} \Big\{ J_1 (1-n_i) \tilde{s}_i^z
- 2 h n_i \tilde{s}_i^x
+\frac{J_1\Delta}{4}(2n_i-1)
\nonumber\\
&&+\left[ 4J_2 n_i n_{i+1} -2(J_2-J_3)(n_i + n_{i+1} - 1) \right]
\tilde{s}_i^x \tilde{s}_{i+1}^x \Big \}.
\end{eqnarray}

It is noteworthy that equation (\ref{new_op}) can be considered
as some nonlinear spin transformation from $s_{1,i}^\alpha$, $s_{2,i}^\alpha$
to new operators ${s'}_{1,i}^{\alpha}$, ${s'}_{2,i}^{\alpha}$,
where e.g. ${s'}_{1,i}^{z}=(n_i-\frac{1}{2})$,
${s'}_{2,i}^\alpha={\tilde s}_{i}^\alpha$.
The transformation can be generated by the following unitary operator:
\begin{eqnarray}
\fl  U=\prod_{i=1}^N
\exp\left[-i\frac{\pi}{2}(s_{1,i}^x+s_{2,i}^x)\right]
\exp\left(i\pi s_{1,i}^x s_{2,i}^x\right)
\exp\left(-i\frac{\pi}{2} s_{2,i}^y\right)
\exp\left(i\pi s_{2,i}^z\right).
\end{eqnarray}
The crucial point is the second factor, which introduces the nonlinearity.
Other terms are the rotation in spin space,
they are used to adjust the transformed operators to the form of (\ref{new_op}).
Finally, the spin operators are transformed as follows:
\begin{eqnarray}
U{s}_{1,i}^x U^+=s_{1,i}^x, 
&U{s}_{1,i}^y U^+=2s_{1,i}^y s_{2,i}^x, 
&U{s}_{1,i}^z U^+=2s_{1,i}^z s_{2,i}^x,
\nonumber\\
U{s}_{2,i}^x U^+=2s_{1,i}^x s_{2,i}^z, 
&U{s}_{2,i}^y U^+=-2s_{1,i}^x s_{2,i}^y, 
&U{s}_{2,i}^z U^+=s_{2,i}^x.
\end{eqnarray}
The transformed Hamiltonian has the form of the quantum Ising chain with composite spins
in an effective longitudinal and transverse magnetic field:
\begin{eqnarray}
U H U^+ &=&\sum_{i=1}^N
\bigg\{
\frac{J_1}{2}\left(1- 2s_{1,i}^z\right)s_{2,i}^z+\frac{J_1 \Delta}{2} s_{1,i}^z
+\Big[J_2(4 s_{1,i}^z s_{1,i+1}^z +1)
\nonumber\\
&+&2J_3( s_{1,i+1}^z + s_{1,i}^z) \Big]
s_{2,i}^x s_{2,i+1}^x
-h(1+2 s_{1,i}^z) s_{2,i}^x
\bigg\}.
\label{gen_ham_spin}
\end{eqnarray}
The straightforward correspondence $s_{1,i}^z=n_i-\frac{1}{2}$,
$s_{2,i}^\alpha=\tilde{s}_{i}^\alpha$ leads to the equivalence between the Hamiltonians
(\ref{gen_ham2}) and (\ref{gen_ham_spin}). It should be noted that such kind of a representation
remains valid also in case of the asymmetric ladder having both diagonal (crossing) Ising interactions
different from each other.

\section{Ground state of the Heisenberg-Ising two-leg ladder}
\label{ground_state}

The Hamiltonian (\ref{gen_ham2}) can also be rewritten in the following more symmetric form:
\begin{eqnarray}
\label{gen_ham3}
 H&{=}&\sum_{i=1}^{N} \Big\{ J_1 (1-n_i) \tilde{s}_i^z
-2h n_i\tilde{s}_i^x
+\frac{J_1\Delta}{4}(2n_i-1)
\nonumber\\
&&{+}\left[ 2(J_2+J_3) n_i n_{i+1} +2(J_2-J_3)(1-n_i)(1-n_{i+1}) \right]
\tilde{s}_i^x \tilde{s}_{i+1}^x \Big\}.
\end{eqnarray}
This form serves in evidence that the effective model splits at $i$th site into two independent chains
provided that two neighbouring bonds are being in different subspaces, i.e. $n_i\neq n_{i+1}$.
The uniform Hamiltonians, when all $n_i=0$ or all $n_i=1$, read:
\begin{eqnarray}
\label{tim}
&&H^0=\sum_{i=1}^N \Big[ J_1\Big(\tilde{s}_i^z-\frac{\Delta}{4}\Big)+2(J_2-J_3)\tilde{s}_i^x \tilde{s}_{i+1}^x \Big],
\\
\label{lim}
&&H^1=\sum_{i=1}^N \Big[ 2(J_2+J_3)\tilde{s}_i^x \tilde{s}_{i+1}^x
-2h\tilde{s}_i^x + \frac{J_1\Delta}{4} \Big].
\end{eqnarray}
If all bonds are in purity states ($n_i=0$),
one obtains the effective Hamiltonian of the Ising chain in the transverse field
that is exactly solvable within Jordan-Wigner fermionization \cite{katsura1963,pfeuty1970}.
If all bonds are in impurity states ($n_i=1$),
one comes to the Ising chain in the longitudinal field
that is solvable by the transfer-matrix method (see e.g. \cite{baxter_book}).
Accordingly, the ground-state  energy of the model in $n_i=0$ subspace is
given by \cite{pfeuty1970}:
\begin{eqnarray}
 e_0^0=\lim_{N\to\infty}\frac{E_0^{0}(N)}{N}
=-\frac{(J_1+|J_2-J_3|)}{\pi}{\mathbf E}(\sqrt{1-\gamma^2})-\frac{J_1\Delta}{4},
\end{eqnarray}
where $\gamma=\frac{J_1-|J_2-J_3|}{J_1+|J_2-J_3|}$ and ${\mathbf E}(\kappa) = \displaystyle \int_0^{\frac{\pi}{2}} \!\!\!\! {\rm d} \theta \sqrt{1 - \kappa^2 \sin^2 \theta}$ is the complete elliptic integral of the second kind.

The ground-state energy of the model in $n_i=1$ subspace is given by
the ground-state energy of the effective Ising chain in the longitudinal field:
\begin{eqnarray}
 e_0^1=\lim_{N\to\infty}\frac{E_0^{1}(N)}{N}
=\left\{
\begin{array}{ll}
 \frac{J_1\Delta}{4}+\frac{J_2+J_3}{2}-|h|, & \mbox{if}\: |h|>(J_2+J_3),\\
 \frac{J_1\Delta}{4}-\frac{J_2+J_3}{2},  & \mbox{if}\: |h|\leq(J_2+J_3).
\end{array}
\right.
\end{eqnarray}
The upper (lower) case corresponds to the ferromagnetically (antiferromagnetically) ordered state
being the ground state at strong (weak) enough magnetic fields.

\subsection{Ground-state phase diagram in a zero field}

If the external field vanishes ($h=0$), it is sufficient to show that the inequality $E_0^0(N_1)+E_0^0(N_2)\geq E_0^0(N_1+N_2)$
holds for any finite transverse Ising chain with free ends
in order to prove that the ground state always corresponds to a uniform bond configuration.
Here $E_0^0(N)$ denotes the ground state energy of the Hamiltonian:
\begin{eqnarray}
\label{finite_tim}
\fl H^0{(N)}&=& {2(J_2-J_3)}\sum_{i=1}^{N-1}\tilde{s}_{i}^x\tilde{s}_{i+1}^x
+ J_1\sum_{i=1}^{N}\Big(\tilde{s}_i^z-\frac{\Delta}{4}\Big).
\end{eqnarray}
Indeed, two independent chains of size $N_1$ and $N_2$ can be represented by the following Hamiltonian:
\begin{eqnarray}
\fl H^0{(N_1,N_2)}&=& {2(J_2-J_3)}\sum_{i=1}^{N_1-1}\tilde{s}_{i}^x\tilde{s}_{i+1}^x
+ 2(J_2-J_3)\sum_{i=N_1+1}^{N_1+N_2-1}\tilde{s}_{i}^x\tilde{s}_{i+1}^x
+ J_1\sum_{i=1}^{N_1+N_2}\Big(\tilde{s}_i^z-\frac{\Delta}{4}\Big)
\nonumber\\
\fl
&=&H^0{(N_1+N_2)} - 2(J_2-J_3)\tilde{s}_{N_1}^x\tilde{s}_{N_1+1}^x.
\end{eqnarray}
If $E_0^0(N)$ and $|\psi_0^{N}\rangle$ are the lowest eigenvalue and eigenstate of $H^0{(N)}$,
then $E_0^0(N_1,N_2)=E_0^0(N_1)+E_0^0(N_2)$ and
$|\psi_0^{N_1,N_2}\rangle=|\psi_0^{N_1}\rangle|\psi_0^{N_2}\rangle$
are the lowest eigenvalue and eigenstate of $H^0{(N_1,N_2)}$.
Now, it is straightforward to show that
\begin{eqnarray}
\label{inequal0}
\fl E_0^0(N_1)+E_0^0(N_2)
&=&\langle\psi_0^{N_1,N_2}| H^0{(N_1+N_2)} |\psi_0^{N_1,N_2}\rangle
\nonumber\\
\fl &-& 2(J_2-J_3)\langle\psi_0^{N_1}|\tilde{s}_{N_1}^x|\psi_0^{N_1}\rangle
\langle\psi_0^{N_2}|\tilde{s}_{N_1+1}^x |\psi_0^{N_2}\rangle
\geq E_0^0(N_1+N_2).
\end{eqnarray}
Here, we have used that $\langle\psi_0^{N_1}|\tilde{s}_{N_1}^x|\psi_0^{N_1}\rangle=0$ for any finite chain \cite{lsm,pfeuty1970},
and that the lowest mean value of the operator $H^0{(N_1+N_2)}$ is achieved in its ground state.

It is easy to show by straightforward calculation
that the same property is valid for the Ising chain with free ends in zero longitudinal field, i.e.
\begin{eqnarray}
\label{inequal1}
\fl E_0^1(N_1+N_2)=
E_0^1(N_1)+E_0^1(N_2)+\frac{|J_2+J_3|}{2}
\leq E_0^1(N_1)+E_0^1(N_2).
\end{eqnarray}

Now, let us prove that the ground state of the whole model may correspond only
to one of uniform bond configurations. It can be readily understood from (\ref{gen_ham3}) that
the effective Hamiltonian does not contain an interaction between spins from two neighbouring bonds if
they are in different subspaces \cite{bose1993}.
It means that the effective model splits into two independent parts
at each boundary ('domain wall') between the purity and impurity states. Thus, the Heisenberg-Ising ladder for
any given configuration of bonds can be considered as a set of independent chains of two kinds and of different sizes.
Then, the ground-state energy of any randomly chosen bond configuration will be as follows:
\begin{eqnarray}
 E&=&E_0^0(N_1)+E_0^0(N_2)+\dots +E_0^1(M_1)+E_0^1(M_2)+\dots
\nonumber\\
&\geq& E_0^0(N_1+N_2+\dots) +  E_0^1(M_1+M_2+\dots)
\label{inq}
\end{eqnarray}
It is quite evident from equations (\ref{inequal0}) and (\ref{inequal1}) that the one uniform
configuration, which corresponds to the state with lower energy than the other one,
must be according to inequality (\ref{inq}) the lowest-energy state (i.e. ground state).
Therefore, the model may show in the ground state the first-order quantum phase transition
when the lowest eigenenergies in both subspaces becomes equal $e_0^1=e_0^0$.
Besides, there also may appear the more striking second-order (continuous) quantum phase transition
at $|J_1|=|J_2-J_3|$  in the ground state, which is inherent to the pure quantum spin chain \cite{pfeuty1970}.
From this perspective, the Heisenberg-Ising ladder can show a variety of quantum phase transitions
in its phase diagram.

\begin{figure}
 \begin{center}
   \includegraphics[width=0.48\textwidth]{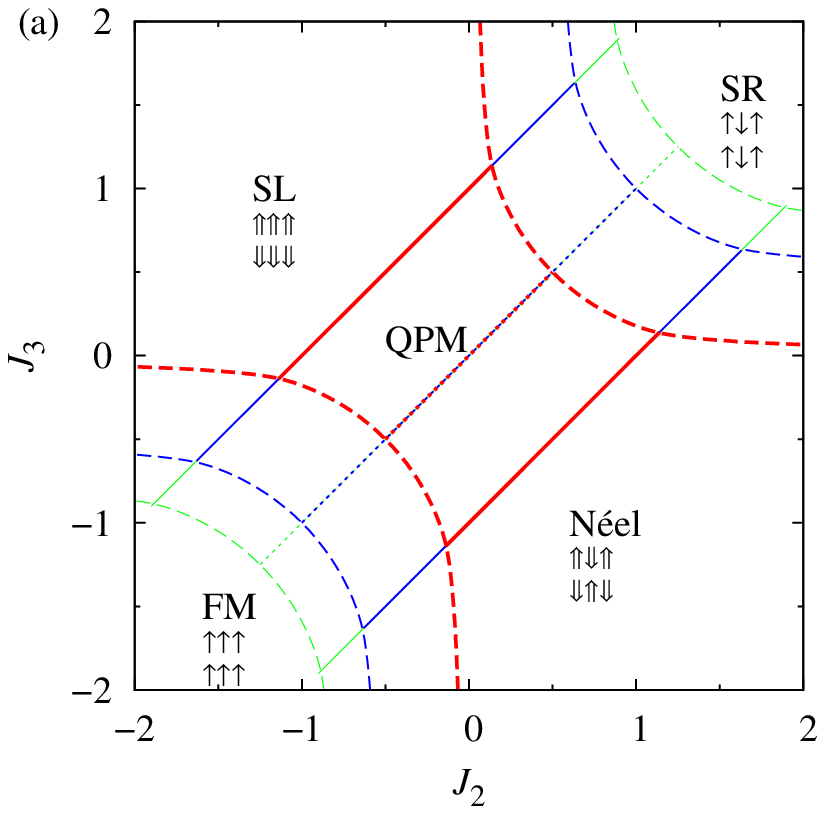}
   \includegraphics[width=0.48\textwidth]{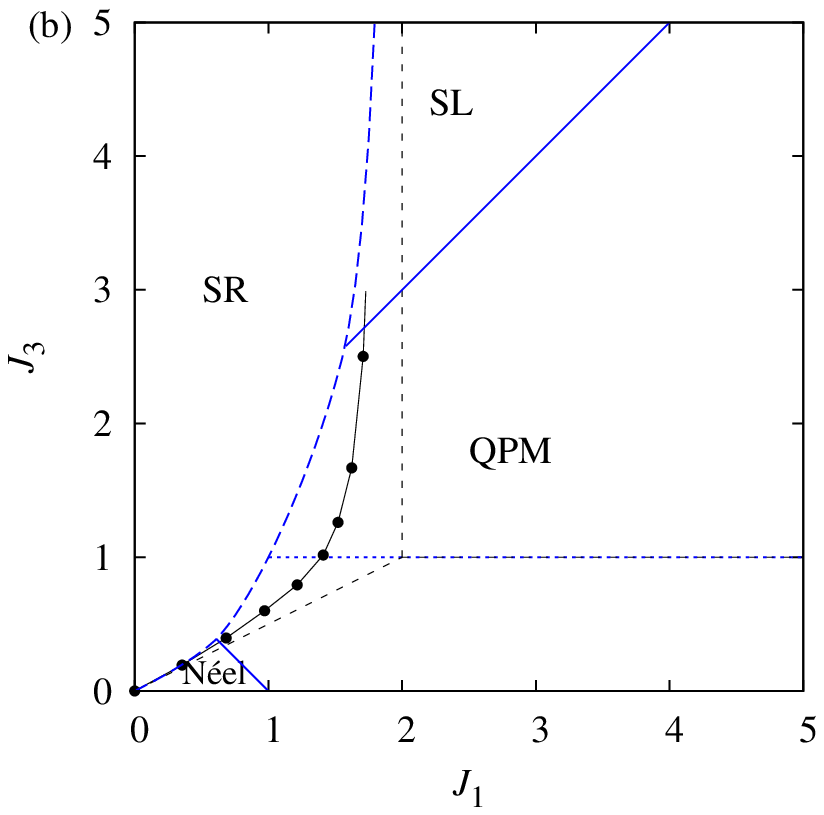}
 \end{center}
\caption{Ground-state phase diagrams for the spin-$\frac{1}{2}$ Heisenberg-Ising two-leg ladder in an absence of the external magnetic field ($h=0$).
Broken (solid) lines denote the lines of first-order (second-order) quantum phase transitions,
dotted lines denote the rung singlet-dimer state. \\
(a) $J_1=1$, $\Delta=0.0$(red), $1.0$(blue), $1.5$(green) (from thick to thin lines);\protect\\
(b) $J_2=1$, $\Delta=1$. Thick solid and broken (blue) lines are the ground-state boundaries of the Heisenberg-Ising ladder, the thin line with dots shows
the ground-state boundary of the Heisenberg ladder \cite{weihong1998} and straight broken lines the ground-state boundaries of the Ising ladder.}
\label{phase_diag1}
\end{figure}

The ground-state phase diagram of the spin-$\frac{1}{2}$ Heisenberg-Ising ladder in a zero magnetic field is shown in figure~\ref{phase_diag1}.
It reflects the symmetry of the model, i.e. it is invariant under the exchange of the Ising interactions
$J_2$ and $J_3$, as well as, under the simultaneous change of the signs of $J_2$ and $J_3$.
Altogether, five different ground states can be recognised in figure~\ref{phase_diag1}:
\begin{enumerate}
 \item {\bf Quantum paramagnetic (QPM) state} for $e_0^0<e_0^1$, and $|J_1|>|J_2-J_3|$:
    the equivalent transverse Ising chain (\ref{tim}) is in the gapped disordered state
    with no spontaneous magnetization $\langle \tilde{s}_i^x\rangle=0$
    and non-zero magnetization $\langle \tilde{s}_i^z\rangle\neq 0$
    induced by the effective transverse field.
    For the initial Heisenberg-Ising ladder it means that the rung singlet dimers
    are dominating on Heisenberg bonds in the ground state. Although each bond may possess the antiferromagnetic order,
    the interaction along legs demolishes it and leads to the disordered quantum paramagnetic state.
 \item {\bf Stripe Leg (SL) state} for $e_0^0<e_0^1$, and $|J_1|<J_3-J_2$:
    the equivalent transverse Ising chain exhibits the spontaneous ferromagnetic ordering
    with $\langle \tilde{s}_i^x\rangle\neq 0$.
    Due to relations (\ref{new_op0}) and (\ref{new_op1}), one obtains for the Heisenberg-Ising ladder
    $\langle s_{1,i}^z\rangle=\langle s_{1,i+1}^z\rangle
     =-\langle s_{2,i}^z\rangle=-\langle s_{2,i+1}^z\rangle\neq 0$.
    This result is taken to mean that the Heisenberg-Ising ladder shows a ferromagnetic order along legs
    and antiferromagnetic order along rungs, i.e. the magnetizations of chains are opposite.
    The staggered magnetization as the relevant
    order parameter in this phase is non-zero and it exhibits
    evident quantum reduction of the magnetization given by:
    $m_{SL}^z=\frac{1}{2N}\sum_{i=1}^N(\langle s_{1,i}^z\rangle-\langle s_{2,i}^z\rangle)
     =\frac{1}{2}[1-J_1^2/(J_2-J_3)^2]^{\frac{1}{8}}$.
     Here we used the result for the spontaneous magnetization of
     transverse Ising chain \cite{pfeuty1970}.
 \item {\bf N\'{e}el state} for $e_0^0<e_0^1$, and $|J_1|<J_2-J_3$:
    the effective transverse Ising chain shows the spontaneous antiferromagnetic order
    with $\langle \tilde{s}_i^x\rangle=(-1)^i m_x\neq 0$. For the Heisenberg-Ising ladder one consequently obtains
    $\langle s_{1,i}^z\rangle=-\langle s_{1,i+1}^z\rangle=-\langle s_{2,i}^z\rangle=\langle s_{2,i+1}^z\rangle\neq 0$.
    Hence, it follows that the nearest-neighbour spins both along legs as well as rungs exhibit
    predominantly antiferromagnetic ordering. The dependence of staggered magnetization as the
    relevant order parameter is quite analogous to the previous case
    $m_{AF}^z=\frac{1}{2N}\sum_{i=1}^N(-1)^{i}(\langle s_{1,i}^z\rangle-\langle s_{2,i}^z\rangle)
     =\frac{1}{2}[1-J_1^2/(J_2-J_3)^2]^{\frac{1}{8}}$.
 \item {\bf Stripe Rung (SR) state} for $e_0^1<e_0^0$, $J_2>0$, $J_3>0$:
    the model shows classical ordering in this phase
    with the antiferromagnetically ordered nearest-neighbour spins along legs
    and the ferromagnetically ordered nearest-neighbour spins along rungs.
 \item {\bf Ferromagnetic (FM) state} for $e_0^1<e_0^0$, $J_2<0$, $J_3<0$:
    the ground state corresponds to the ideal fully polarized ferromagnetic spin state.
\end{enumerate}

\begin{figure}
 \begin{center}
   \includegraphics[width=0.48\textwidth]{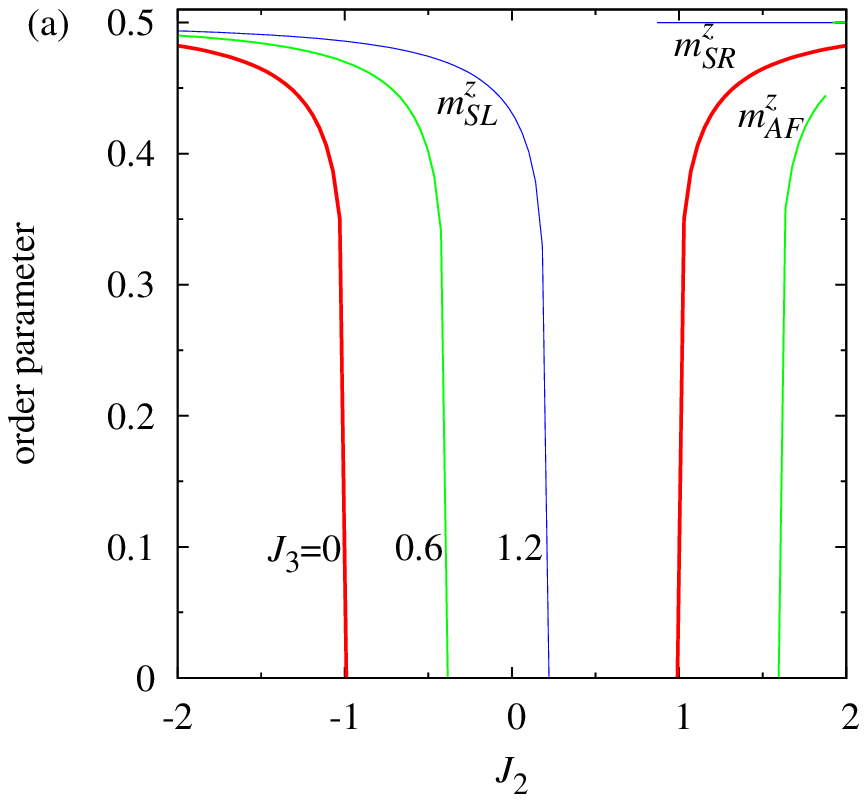}
   \includegraphics[width=0.48\textwidth]{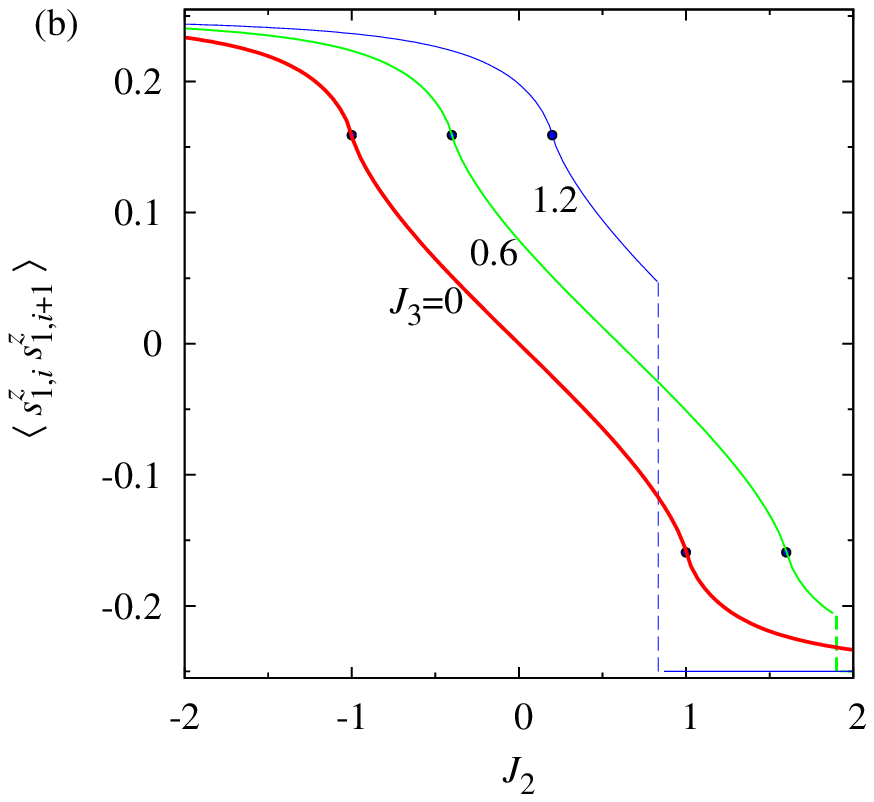}
 \end{center}
\caption{Zero-temperature variations of: (a) the order parameters; (b) the nearest-neighbour spin-spin correlation along legs;
as a function of the Ising intra-leg interaction $J_2$ for $J_1=1$, $\Delta=1$ and three different values of the crossing Ising
interaction $J_3=0,0.6,1.2$.}
\label{correlation}
\end{figure}

The results displayed in figure~\ref{phase_diag1}(a) demonstrate that the Heisenberg-Ising ladder is in the disordered QPM phase whenever
a relative strength of both Ising interactions $J_2$ and $J_3$ is sufficiently small compared to the Heisenberg intra-rung interaction $J_1$.
It is noteworthy, moreover, that the QPM phase reduces to a set of fully non-correlated singlet dimers placed on all rungs (the so-called
rung singlet-dimer state) along the special line $J_2=J_3$ up to $e_0^0<e_0^1$, which is depicted in figure~\ref{phase_diag1} by dotted lines.
Under this special condition, the intra-rung spin-spin correlation represents the only non-zero pair correlation function and all the other
short-ranged spin-spin correlations vanish and/or change their sign across the special line $J_2=J_3$. It should be remarked that the completely
identical ground state can also be found in the symmetric Heisenberg two-leg ladder with $J_2=J_3$ (see figure~\ref{phase_diag1}(b)). To compare with,
the rung singlet-dimer state is being the exact ground state of the symmetric Heisenberg-Ising ladder with $\Delta=1$ for $J_1>J_2$, while the symmetric Heisenberg
ladder displays this simple factorizable ground state for $J_1>1.4015 J_2$ \cite{gelfand1991,xian1995} (this horizontal line is for clarity not shown
in figure~\ref{phase_diag1}(b) as it exactly coincides with the ground-state boundary of the Heisenberg-Ising ladder extended over larger parameter space).
It should be stressed, however, that the short-range spin-spin correlations become non-zero in QPM whenever $J_2 \neq J_3$ even if this
phase still preserves its disordered nature with the prevailing character of the rung singlet-dimer state. To support this statement, the
zero-temperature dependencies of the order parameters and the nearest-neighbour spin-spin correlation along legs are plotted in figure~\ref{correlation}.
The relevant order parameters evidently disappear in QPM, whereas the nearest-neighbour correlation function changes its sign when passing
through the special line $J_2=J_3$ of the rung singlet-dimer state.

Note furthermore that the Heisenberg-Ising ladder undergoes the second-order quantum phase transition from the disordered
QPM phase to the spontaneously long-range ordered N\'{e}el or SL phase, which predominantly appear in the parameter region
where one from both Ising couplings $J_2$ and $J_3$ is being antiferromagnetic and the other one is ferromagnetic.
The significant quantum reduction of staggered order parameters implies an obvious quantum character of both
N\'{e}el as well as SL phases in which the antiferromagnetic correlation on rungs still dominates (see figure~\ref{correlation}).
Thus, the main difference between both the quantum long-range ordered phases emerges in the character nearest-neighbour
correlation along legs, which is ferromagnetic in the SL phase but antiferromagnetic in the N\'{e}el phase.
The continuous vanishing of the order parameters depicted in figure~\ref{correlation}(a) provides a direct evidence of
the second-order quantum phase transition between the disordered QPM phase and the ordered SL (or N\'{e}el) phase,
which is also accompanied with the weak-singular behaviour of the nearest-neighbour correlation function visualized
in figure~\ref{correlation}(b) by black dots.
Here it should be noted that the $zz$ correlation function of the Heisenberg-Ising ladder
can be easily derived from the result of
the $xx$ correlation function of the transverse Ising chain calculated in \cite{pfeuty1970}.
Last but not least, the strong enough Ising interactions $J_2$ and $J_3$ may
break the antiferromagnetic correlation along rungs and lead to a presence of the fully ordered ferromagnetic rung states FM or SR
depending on whether both Ising interactions are ferromagnetic or antiferromagnetic, respectively. It should be noticed
that this change is accompanied with a discontinuous (first-order) quantum phase transition on behalf of a classical character of
both FM and SR phases, which can be also clearly seen in figure~\ref{correlation} from an abrupt change of the order parameters
as well as the nearest-neighbour correlation function.

Finally, let us also provide a more detailed comparison between the ground-state phase diagrams of the Heisenberg, Heisenberg-Ising
and Ising two-leg ladders all depicted in figure~\ref{phase_diag1}(b). One can notice that the displayed ground-state phase diagrams of the
Heisenberg and Heisenberg-Ising two-leg ladders have several similar features. The phase boundary between the classically ordered
SR phase and three quantum phases (QPM, SL, N\'{e}el) of the Heisenberg-Ising ladder follows quite closely the first-order phase boundary
between the Haldane-type phase and the dimerized phase of the pure Heisenberg ladder obtained using the series expansion and exact diagonalization \cite{weihong1998}.
Of course, the most fundamental difference lies in the character of SR and Haldane phases, because the former phase exhibits a classical long-range
order contrary to a more spectacular topological order of the pure quantum Haldane-type phase with a non-zero string order parameter \cite{kim2008}
even though the ferromagnetic intra-rung correlation is common for both phases. The difference between the ground states of the Heisenberg-Ising and
pure Heisenberg two-leg ladder becomes much less pronounced in the parameter region with the strong intra-rung interaction $J_1$, which evokes the
quantum phases in the ground state of both these models. In the case of a sufficiently strong frustration $J_2\sim J_3$, the ground state of the
Heisenberg-Ising ladder forms the disordered QPM phase, whereas the antiferromagnetic or ferromagnetic (N\'eel or SL) long-range order emerges
along the legs if the diagonal coupling $J_3$ is much weaker or stronger than the intra-leg interaction $J_2$. Note that the quantum N\'eel and
SL phases have several common features (e.g. predominant antiferromagnetic correlations along rungs) with the disordered QPM phase from which
they evolve when crossing continuous quantum phase transitions given by the set of equations: $J_3=J_2-J_1$ (for $J_3<J_2$) and $J_3=J_2+J_1$ (for $J_3>J_2$).
It cannot be definitely ruled out whether or not these quantum ordered phases may become the ground state of the pure Heisenberg ladder,
because they are also predicted by the bond-mean-field approximation \cite{ramakko2007,azzouz2008,ramakko2008} but have not been found
by the most of the numerical methods. Further numerical investigations of the Heisenberg ladder are therefore needed to clarify this unresolved issue.

\subsection{Ground-state phase diagram in a non-zero field}

When the external field $h$ is switched on, the inequality for the ground states of the Ising chain in the longitudinal field (\ref{inequal1}) is generally valid only for $J_1+J_2\leq 0$ or $J_1+J_2\geq 2|h|$. Hence, it is necessary to modify the procedure
of finding the ground states inside the region where the relation (\ref{inequal1}) is broken. However, one may use with a success the method suggested by Shastry and Sutherland \cite{shastry1981,chandra2006} in order to find the ground states inside this parameter region.
Let us represent our effective Hamiltonian (\ref{gen_ham3}) in the form:
\begin{eqnarray}
&& H=\sum_{i=1}^N H_{i,i+1},
\nonumber\\
&& H_{i,i+1}=
\frac{1}{2}\sum_{l=i}^{i+1} \Big\{ J_1 (1-n_l) \tilde{s}_l^z
-2h n_l\tilde{s}_l^x
+\frac{J_1\Delta}{4}(2n_l-1) \Big\}
\nonumber\\
&& +\left[ 2(J_2+J_3) n_i n_{i+1} +2(J_2-J_3)(1-n_i)(1-n_{i+1}) \right]
\tilde{s}_i^x \tilde{s}_{i+1}^x.
\end{eqnarray}
Then, one can employ the variational principle  implying that $E_0\geq \sum_{l=1}^N E_0(l,l+1)$,
where $E_0(l,l+1)$ is the lowest eigenenergy of $H_{l,l+1}$.

Looking for the eigenenergies of $H_{l,l+1}$ it is enough
to find the lowest eigenstate of each bond configuration:
\begin{itemize}
 \item $n_i=n_{i+1}=0$,
$E_0^{0,0}(i,i+1)=-\frac{1}{2}\sqrt{J_1^2+(J_2-J_3)^2}-\frac{J_1\Delta}{4}$;

 \item $n_i=n_{i+1}=1$,
\begin{eqnarray}
 E_0^{1,1}(i,i+1)=\left\{
\begin{array}{ll}
 \frac{J_1\Delta}{4}+\frac{J_2+J_3}{2}-|h|, & \mbox{if}\: |h|>(J_2+J_3) (\rm\bf{FM}),\\
 \frac{J_1\Delta}{4}-\frac{J_2+J_3}{2},  & \mbox{if}\: |h|\leq(J_2+J_3) (\rm\bf{AF}).
\end{array}
\right.
\nonumber
\end{eqnarray}

 \item $n_i=0$, $n_{i+1}=1$ ($n_i=1$, $n_{i+1}=0$),
  $E_0^{0,1}(i,i+1)=E_0^{1,0}(i,i+1)=-\frac{J_1}{4}-\frac{|h|}{2}$.

\end{itemize}
The phase corresponding to the alternating bond configuration $n_i=0$, $n_{i+1}=1$
will be hereafter referred to as the staggered bond (SB) phase. It should be mentioned that
there does not exist in the SB phase any correlations between spins from different rungs
and the overall energy comes from the intra-rung spin-spin interactions and the Zeeman's energy
of the fully polarized rungs \cite{bose1993}.
It can be easily seen from a comparison of $E_0^{0,1}(i,i+1)$ and $E_0^{1,1}(i,i+1)$
that the eigenenergy of the SB phase $E_0^{0,1}(i,i+1)$ has always lower energy
than the lowest eigenenergy of the fully polarized state $E_0^{1,1}(i,i+1)$ inside the stripe:
\begin{eqnarray}
\label{sb_cond1}
 J_2+J_3 - \frac{J_1(1+\Delta)}{2}\leq |h| \leq J_2+J_3 + \frac{J_1(1+\Delta)}{2}.
\end{eqnarray}
Note that the ferromagnetic ordering is preferred for external fields above this stripe
$|h| \geq J_2+J_3 + \frac{J_1(1+\Delta)}{2}$, while the antiferromagnetic ordering becomes
the lowest-energy state below this stripe $|h| \leq J_2+J_3 - \frac{J_1(1+\Delta)}{2}$ .

If one compares the respective eigenenergies of the staggered bond phase $E_0^{0,1}(i,i+1)$
and the uniform purity phase $E_0^{0,0}(i,i+1)$ ($n_i=n_{i+1}=0$), one gets another condition
implying that $E_0^{0,1}(i,i+1)$ becomes lower than $E_0^{0,0}(i,i+1)$ only if:
\begin{eqnarray}
\label{sb_cond2}
 |h|\geq \sqrt{J_1^2+(J_2-J_3)^2}-\frac{J_1(1-\Delta)}{2}.
\end{eqnarray}
This means that the SB phase with the overall energy $E_0^{0,1}=-N(J_1+2|h|)/4$
becomes the ground state inside the region confined by conditions (\ref{sb_cond1}) and (\ref{sb_cond2}).
The lowest field, which makes the SB phase favourable, can be also found as:
\begin{eqnarray}
 |h_{min}|=\frac{J_1(1+\Delta)}{2}.
\end{eqnarray}

Similarly, one may also find the condition under which two eigenenergies corresponding to the uniform impurity configuration
$E_0^{1,1} (n_i=n_{i+1}=1)$ become lower than the respective eigenenergy of the uniform purity configuration $E_0^{0,0} (n_i=n_{i+1}=0)$. The ferromagnetic state of the uniform impurity configuration becomes lower if the external field exceeds the boundary value:
\begin{eqnarray}
\label{fm_cond2}
 |h| \geq \frac{1}{2}\left[
J_2+J_3+J_1\Delta + \sqrt{J_1^2+(J_2-J_3)^2}
\right],
\end{eqnarray}
whereas the condition for the antiferromagnetic state of the uniform impurity configuration is independent of the external field:
\begin{eqnarray}
\label{af_cond2}
 J_1\Delta+\sqrt{J_1^2+(J_2-J_3)^2}\leq (J_2+J_3).
\end{eqnarray}

It is worthy of notice that it is not possible to find the ground state outside the boundaries (\ref{sb_cond2}), (\ref{fm_cond2}), (\ref{af_cond2}) using the variational principle. However, it is shown in the appendix that the bond configuration, which corresponds to the ground state, cannot exceed period two. Therefore, one has to search for the ground state just among the states that correspond to the following bond configurations: all $n_i=0$; all $n_i=1$; $n_{2i-1}=0$, $n_{2i}=1$ ($n_{2i-1}=1$, $n_{2i}=0$). In this respect, two phases are possible inside the stripe given by (\ref{sb_cond1}). The ground state of the Heisenberg-Ising ladder forms the SB phase if $e_0^{0,1}<e_0^{0}$:
\begin{eqnarray}
 |h| \geq \frac{2(J_1+|J_2-J_3|)}{\pi}{\mathbf E}(\sqrt{1-\gamma^2})-\frac{J_1(1-\Delta)}{2}.
\end{eqnarray}

\begin{figure}
 \begin{center}
   \includegraphics[width=0.48\textwidth]{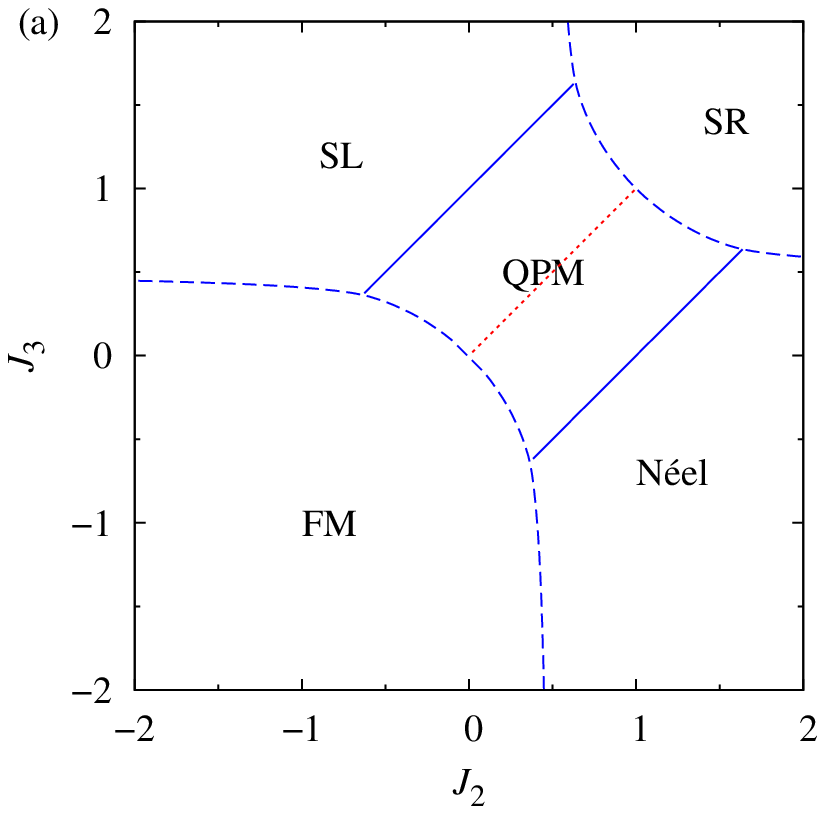}
   \includegraphics[width=0.48\textwidth]{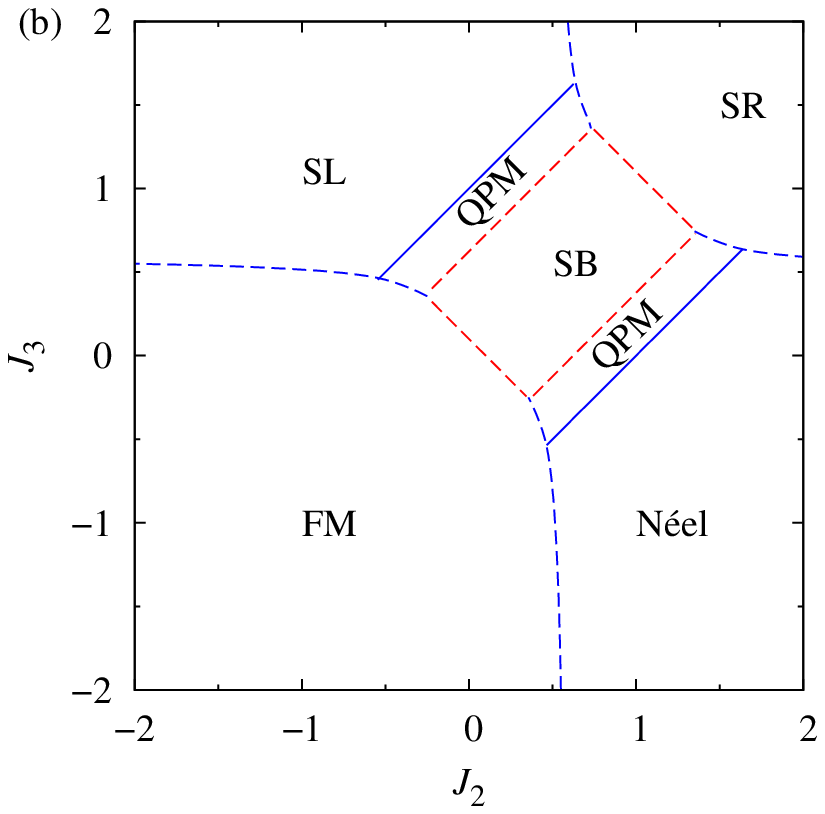}
 \end{center}
\caption{ Ground-state phase diagrams for the Heisenberg-Ising two-leg ladder
in the $J_2-J_3$ plane for $J_1=1$, $\Delta=1$ and two different values of the external field:
(a) $h=1.0$; (b) $h=1.1$.}
\label{phase_diag3}
\end{figure}
\begin{figure}
 \begin{center}
   \includegraphics[width=0.48\textwidth]{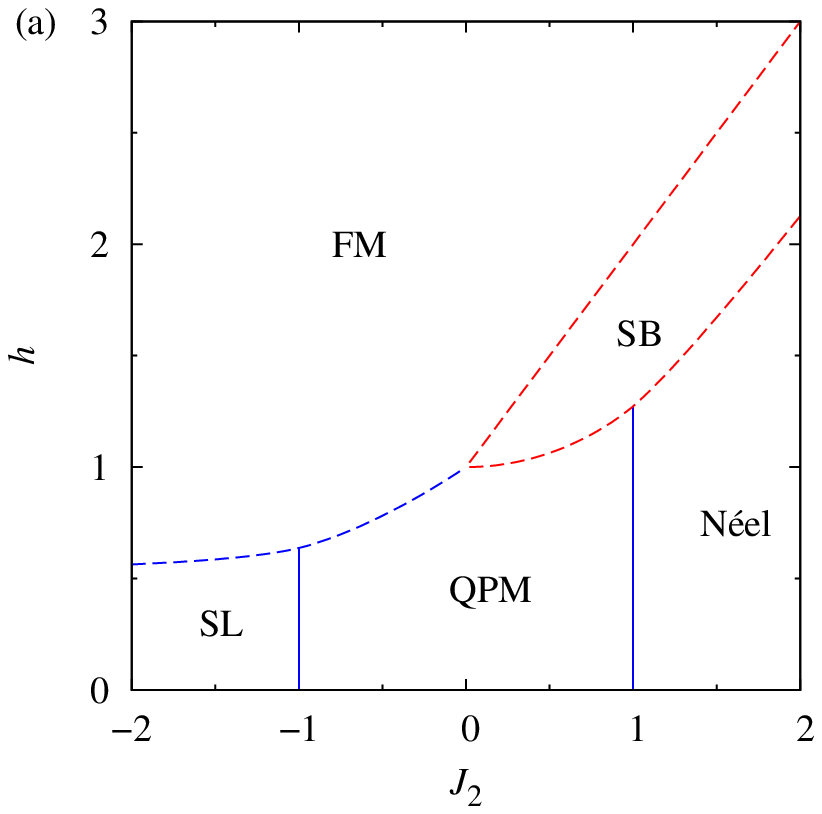}
   \includegraphics[width=0.48\textwidth]{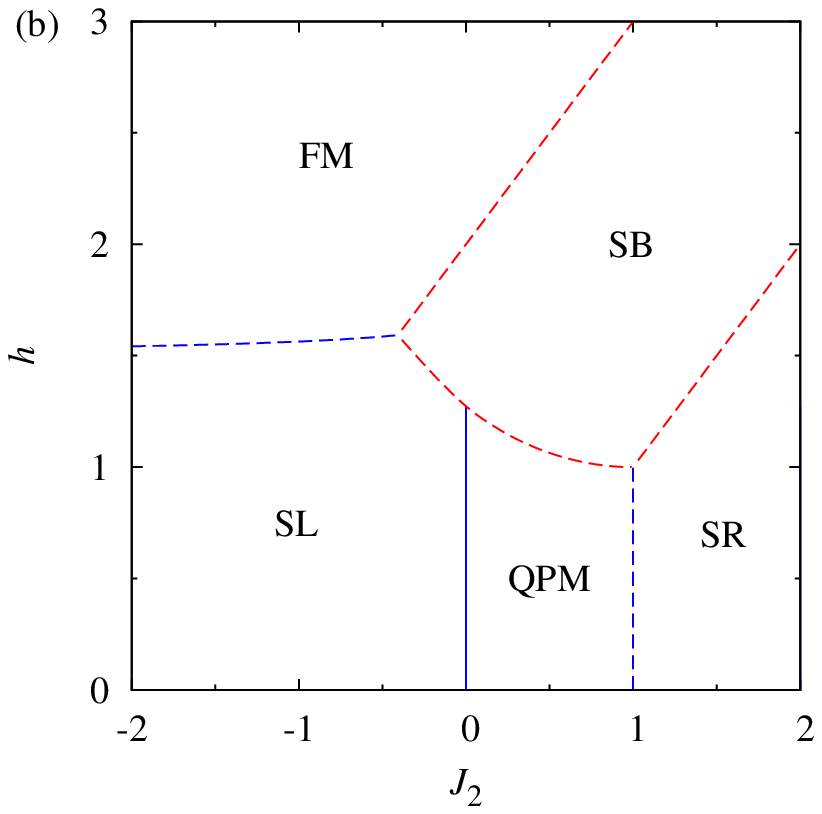}
 \end{center}
\caption{ Ground-state phase diagrams for the Heisenberg-Ising two-leg ladder
in the $J_2-h$ plane for $J_1=1$, $\Delta=1$ and two different values of the
crossing interaction: (a) $J_3=0$; (b) $J_3=1$.}
\label{phase_diag4}
\end{figure}
\begin{figure}
 \begin{center}
   \includegraphics[width=0.48\textwidth]{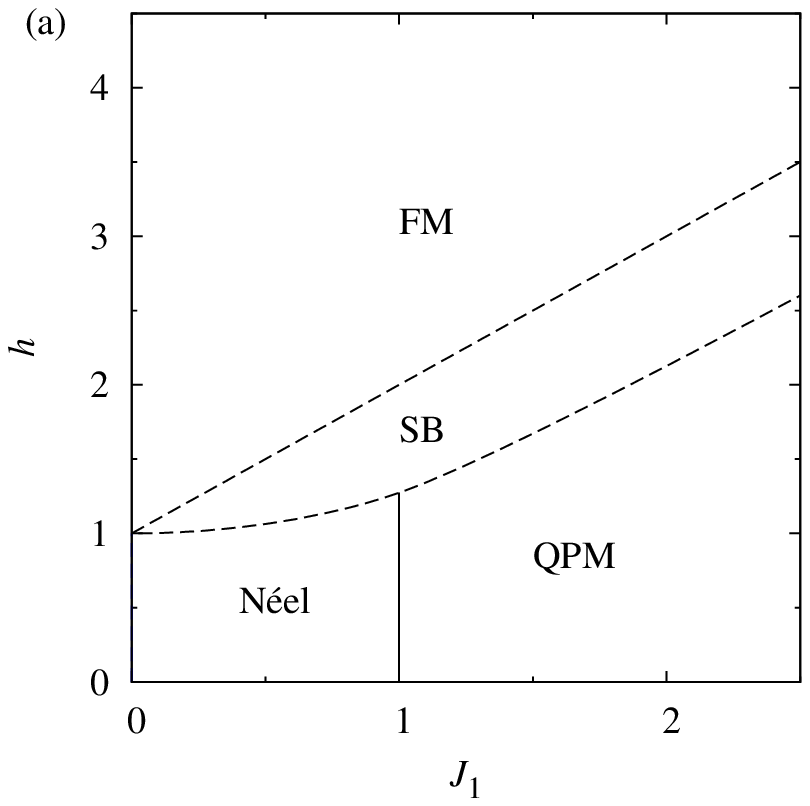}
   \includegraphics[width=0.48\textwidth]{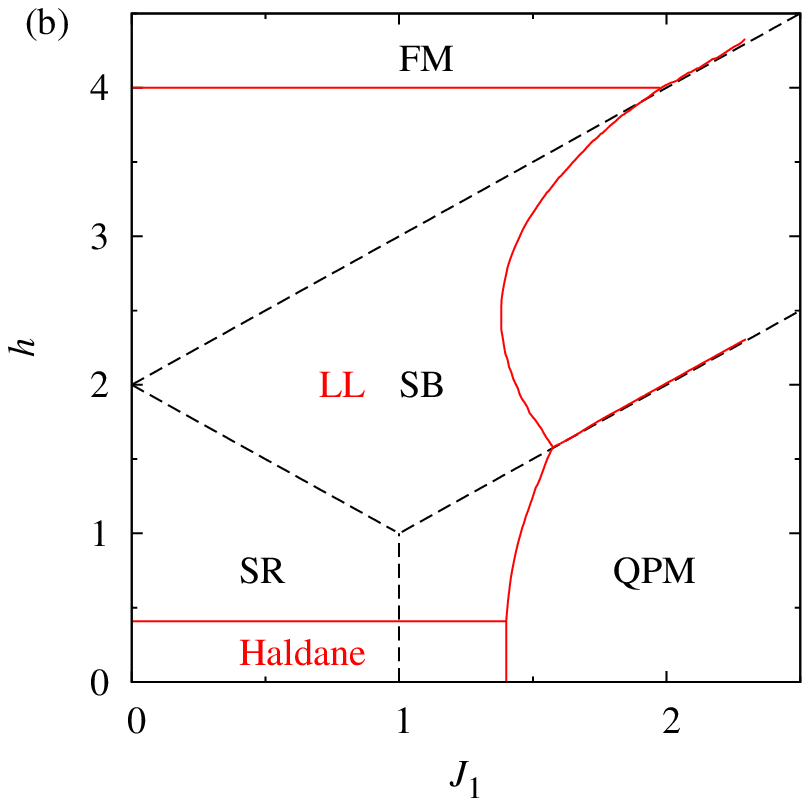}
 \end{center}
\caption{ Ground-state phase diagram for the Heisenberg-Ising two-leg ladder
in the $J_1-h$ plane for $J_2=1$, $\Delta=1$
and two different values of the crossing interaction: (a) $J_3=0$; (b) $J_3=1$.
For a comparison, figure~\ref{field}(b) depicts by red lines the ground-state phase diagram
of the pure Heisenberg two-leg ladder adapted from \cite{honecker2000}.}
\label{field}
\end{figure}

Several ground-state phase diagrams are plotted
in figures~\ref{phase_diag3}-\ref{field} for a non-zero magnetic field.
The most interesting feature stemming from these phase diagrams is
that the SB phase may become the ground state
for the magnetic field $h\geq \frac{J_1(1+\Delta)}{2}$,
which is sufficiently strong to break the rung singlet-dimer state.
It can be observed from figure~\ref{phase_diag3} that the SB phase indeed
evolves along the line of the rung singlet-dimer state
and replaces the QPM phase in the ground-state phase diagram.
The external magnetic field may thus cause an appearance
of another peculiar quantum SB phase with the translationally
broken symmetry, i.e. the alternating singlet and fully polarized triplet bonds
on the rungs of the two-leg ladder. Hence, it follows that the SB phase emerges
at moderate values of the external magnetic field and it consequently leads to a presence
of the intermediate magnetization plateau at a half of the saturation magnetization.
It is quite apparent from figures~\ref{phase_diag4}(a), \ref{field}(a)
that the Heisenberg-Ising ladder without the frustrating
diagonal Ising interaction $J_3=0$ exhibits this striking magnetization plateau just for
the particular case of the antiferromagnetic Ising intra-leg interaction $J_2>0$.
On the other hand, the fractional magnetization plateau inherent to a presence of the SB phase
is substantially stabilized by the spin frustration triggered by the non-zero diagonal
Ising interaction $J_3 \neq 0$ and hence, the plateau region may even extend over a relatively narrow
region of the ferromagnetic Ising intra-leg interaction $J_2<0$ as well (see figure~\ref{phase_diag4}(b)).

It should be noted that the same mechanism for an appearance of the magnetization plateau
has also been predicted for the spin-$\frac{1}{2}$ Heisenberg two-leg ladder
by making use of exact numerical
diagonalization and DMRG methods \cite{honecker2000,okazaki2000,sakai2000,chandra2006,michaud2010}.
Let us therefore conclude our study by comparing the respective ground-state phase diagrams
of the Heisenberg-Ising and Heisenberg two-leg ladders in a presence of the external magnetic field
displayed in figure~\ref{field}.
According to this plot, both models give essentially the same magnetization
process in the strong-coupling limit of the Heisenberg intra-rung interaction $J_1 \gtrapprox 1.5$,
where two subsequent field-induced transitions in the following order QPM-SB-FM
can be observed and consequently, the magnetization exhibits two abrupt jumps at
the relevant transition fields. On the other hand, the fundamental differences can
be detected in the relevant magnetization process of the Heisenberg-Ising and Heisenberg ladders
in the weak-coupling limit of the intra-rung interaction $J_1$. At low fields, the quantum
Haldane-like phase constitutes the ground state of the pure Heisenberg ladder in contrast
to the classical SR phase, which is being the low-field ground state of the Heisenberg-Ising ladder.
In addition, the Heisenberg-Ising ladder still exhibits a magnetization plateau corresponding to
a gapped SB phase at intermediate values of the magnetic field in this parameter space,
while a continuous increase of the magnetization can be observed in the pure Heisenberg
ladder due to a presence of the gapless Luttinger-liquid phase at moderate fields. Finally,
it is also noteworthy that the saturation fields towards FM phase are also quite different
for the Heisenberg-Ising and Heisenberg ladders in the weak-coupling limit of the Heisenberg
intra-rung interaction.

\section{Conclusions}
\label{conclusions}
The frustrated Heisenberg-Ising two-leg ladder was considered
within the rigorous approach
using that $z$-projection of total spin on a rung is a conserved quantity.
By means of the pseudospin representation of bond states, we have proved
the exact mapping correspondence between the investigated model and some generalized spin-$\frac{1}{2}$
quantum Ising chain with the composite spins in an effective longitudinal and transverse field.
We have also found the unitary transformation  which reproduces the analogous mapping between the models.
While the bond-state representation has more transparent physical interpretation,
the unitary transformation gives the complete relation between spin operators of both models,
and it might be useful for searching quantum spin ladders which admit exact solutions.

It has been
shown that the ground state of the model under investigation must correspond to a regular bond configuration
not exceeding the period two. Hence, the true ground state will be the lowest eigenstate of either the transverse
Ising chain, the Ising chain in the longitudinal field, or the non-interacting spin-chain model in the
staggered longitudinal-transverse field.

The most interesting results to emerge from the present study are closely related to an extraordinary diversity of
the constructed ground-state phase diagrams and the quantum phase transitions between different ground-state phases.
In an absence of the external magnetic field, the ground-state phase diagram constitute four different ordered phases
and one quantum paramagnetic phase. The disordered phase was characterized through short-range spin correlations,
which indicate a dominating character of the rung singlet-dimer state in this phase. On the other hand, the order
parameters have been exactly calculated for all the four ordered phases, two of them having classical character
and two purely quantum character as evidenced by the quantum reduction of the staggered magnetization in the latter
two phases. Last but not least, it has been demonstrated that the external magnetic field of a moderate strength may cause
an appearance of the peculiar SB phase with alternating character of singlet and triplet bonds on the rungs of two-leg ladder.
This latter finding is consistent with a presence of the fractional magnetization plateau at a half
of the saturation magnetization in the relevant magnetization process.

\ack
The authors are grateful to Oleg Derzhko for several useful comments and suggestions.
T.V. was supported by the National Scholarship Programme of the Slovak Republic
for the Support of Mobility of Students, PhD Students, University Teachers and Researchers.
J.S. acknowledges the financial support under the grant VEGA 1/0234/12.

\appendix\section{Ground-state bond configuration in magnetic field}
\setcounter{section}{1}

In general each state of the model can be represented as an array of alternating purity and impurity
non-interacting clusters of different length.
One can formally write the lowest energy of some configuration as
\begin{eqnarray}
 E=E_0(N_1,M_1)+E_0(N_2,M_2)+\dots +E_0(N_L,M_L),
\end{eqnarray}
where $E_0(N_i,M_i)=E_0^0(N_i)+E_0^1(M_i)$ is the lowest energy of
the complex of two independent spin chains which correspond to the purity and impurity bond states,
$N_1+M_1+N_2+M_2+\dots N_L+M_L=N$.
If $e_0(N_i,M_i)=E_0(N_i,M_i)/(N_i+M_i)$ corresponds to the lowest energy per spin among other clusters,
it is evident that the ground state configuration is alternating purity and impurity clusters
of length $N_i$ and $M_i$.

Let us consider at first the case when $h$ is restricted by condition (\ref{sb_cond1}).
If the configuration contains the cluster of more than 2 impurity bonds,
its energy can be lowered by adding pure bond in-between. If condition (\ref{sb_cond1}) is valid,
such a configuration can achieve a lower energy. Using this procedure we can reduce the number of sites
in impurity clusters to 2, i.e. only $E_0(N,1)$ or $E_0(N,2)$ can correspond to the ground state.

\begin{figure}
 \begin{center}
   \includegraphics[width=0.48\textwidth]{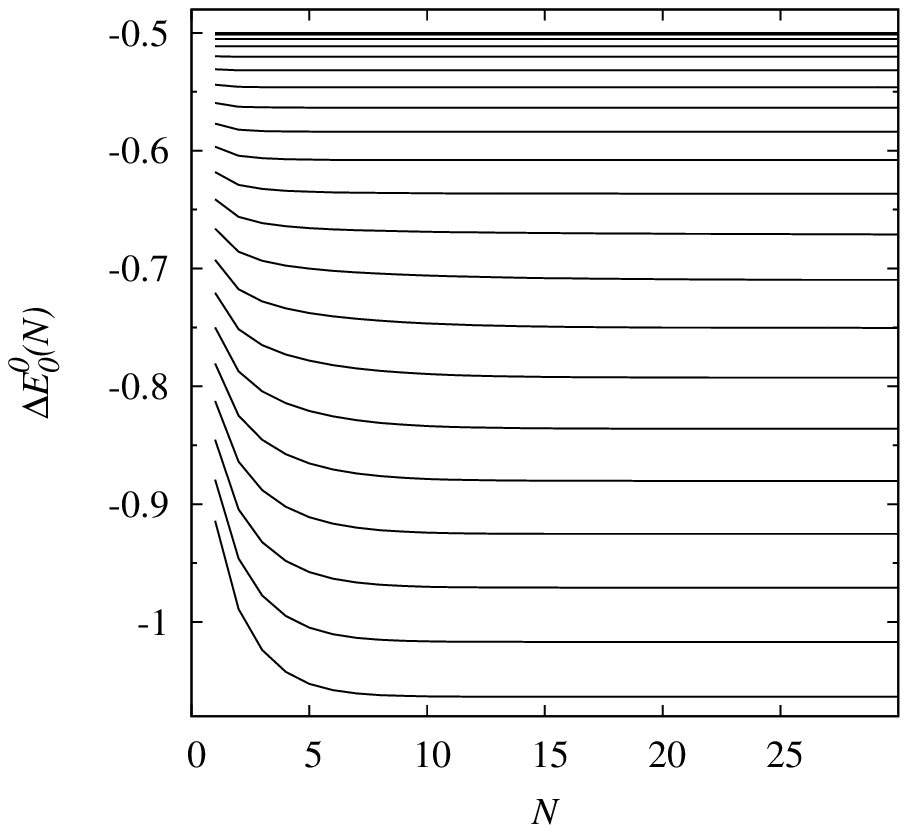}
 \end{center}
\caption{$\Delta E_0^0(N)=E_0^0(N+1)-E_0^0(N)$ as a function of system size $N$
for the transverse Ising chain with free ends (\ref{finite_tim}): $J_1=1$, $\Delta=0$,
$|J_2-J_3|=0,0.2,0.4,\dots,4$ from top to bottom.}
\label{de0_3d}
\end{figure}

Define the energy per spin for each configuration as:
\begin{eqnarray}
 e_0(N,1)=\frac{E_0^0(N)+E_0^1(1)}{N+1},
\\
 e_0(N,2)=\frac{E_0^0(N)+E_0^1(2)}{N+2}.
\end{eqnarray}
If $e_0(1,1)<e_0(2,1)$ and so on ($e_0(N,1)<e_0(N+1,1)$) the ground state corresponds to the staggered bond configuration.
Suppose that for some $N$
\begin{eqnarray}
\fl \Delta e_0(N-1)&=&e_0(N,1)-e_0(N-1,1)
\nonumber\\
\fl &=&\frac{1}{N(N+1)}(NE_0^0(N)-(N+1)E_0^0(N-1)-E_0^1(1))<0.
\label{cond1}
\end{eqnarray}
Consequently, it is clear that
\begin{eqnarray}
\fl  \Delta e_0(N)
&=&\frac{1}{(N+1)(N+2)}((N+1)E_0^0(N+1)-(N+2)E_0^0(N)-E_0^1(1))<0,
\label{cond2}
\end{eqnarray}
if $E_0^0(N+1)-2E_0^0(N)+E_0^0(N-1)\leq 0$, i.e.
the ground state energy of finite transverse Ising chain is
a convex function of $N$.
The last condition is ensured by non-increasing
$\Delta E_0^0(N)=E_0^0(N+1)-E_0^0(N)$ ($\Delta E_0^0(N)\geq\Delta E_0^0(N+1)$)
that can be shown by numerical calculations for finite chains
(see figure~\ref{de0_3d}) using the numerical approach \cite{derzhko1997}.
For $J_2=J_3$ we get the free spin Hamiltonian
and $E_0^0(N)$ depends linearly on $N$.
Conditions (\ref{cond1}),(\ref{cond2}) together mean that $e_0(N,1)$ may have only one extremum and it is the maximum.
Therefore, $e_0(N,1)$ may take the minimal value only in two limiting cases $N=1$, $N\to\infty$.
The same result can be analogously obtained for $e_0(N,2)$.
Now we can compare the energies of $(N,1)$ and $(N,2)$ configurations.
For $|h|\ge\frac{J_1+J_2}{2}$
\begin{eqnarray}
 e_0(1,1)-e_0(1,2)=\frac{1}{6}
\left(-\frac{J_1(1+\Delta)}{2}-(J_2+J_3)+|h|
\right)<0,
\end{eqnarray}
due to the upper boundary of condition (\ref{sb_cond1}).
To summarize, the ground state configuration in the region confined by (\ref{sb_cond1})
can correspond to either staggered bond or uniform purity configuration.

Let us consider the ground state for $|h|>J_2+J_3+\frac{J_1(1+\Delta)}{2}$.
Similarly to the previous case we can prove that $e_0(N,M)$ is a convex function
of $N$ and $M$ if the ground state energies of the uniform chains
have the following properties:
\begin{eqnarray}
E_0^0(N+1)-2E_0^0(N)+E_0^0(N-1)\leq 0,
\nonumber\\
E_0^1(N+1)-2E_0^1(N)+E_0^1(N-1)\leq 0.
\end{eqnarray}
One can find that $E_0^1(N+1)-E_0^1(N)=\frac{|J_2-J_3|}{4}$,
and $E_0^1(N+1)-2E_0^1(N)+E_0^1(N-1)= 0$.
That is why the minimal value of the ground state can be achieved
in one of three cases: staggered bond, purity and impurity configuration.
The staggered bond configuration should be excluded from this list
for $|h|>J_2+J_3+\frac{J_1(1+\Delta)}{2}$,
since it cannot be the ground state due to the variational principle.

\end{document}